\documentclass[%
 reprint,
superscriptaddress,
nofootinbib,
 amsmath,amssymb,
 aps,
showkeys
]{revtex4-2}

\usepackage{graphicx}
\usepackage{dcolumn}
\usepackage{bm}
\usepackage{hyperref}
\usepackage[mathlines]{lineno}
\usepackage{fontenc}
\usepackage{graphicx}
\usepackage{dcolumn}
\usepackage{bm}
\usepackage{booktabs}
\usepackage{array,multirow}
\usepackage{hyperref}  
\usepackage{amsmath}
\usepackage{amsfonts}
\usepackage{amssymb}
\usepackage{mathrsfs}
\usepackage{enumerate}
\usepackage{fancyhdr}
\usepackage{xcolor}
\usepackage{graphicx}
\usepackage{listings}
\usepackage{hyperref}
\usepackage{float}
\usepackage{verbatim}
\usepackage{braket}
\usepackage{centernot}
\usepackage{setspace}
\usepackage{endnotes}
\usepackage[normalem]{ulem}
\usepackage{wrapfig}
\usepackage{multirow}
\usepackage{comment}
\usepackage{kantlipsum}
\allowdisplaybreaks
\usepackage{lipsum, babel}
\DeclareMathOperator{\sgn}{sgn}

\hypersetup{breaklinks = true, colorlinks = true, citecolor = blue, linkcolor = blue, urlcolor = blue}
\usepackage{cleveref}
\newcommand{\be}{\begin{equation}}
\newcommand{\ee}{\end{equation}}
\newcommand{\bea}{\begin{eqnarray}}
\newcommand{\eea}{\end{eqnarray}}

\newcommand{\Tr}{\textrm{Tr}}

\begin{document}


\title{Measurement of Bell-type inequalities and quantum entanglement \\from $\Lambda$-hyperon spin correlations at high energy colliders}

\author{Wenjie~Gong}
\email{wenjiegong@college.harvard.edu}
\affiliation{%
 Department of Physics, Harvard University, Cambridge, Massachusetts 02138, USA
}%

\author{Ganesh Parida}
\email{parida@wisc.edu}
\affiliation{%
 Department of Physics, University of Wisconsin-Madison, Madison, Wisconsin 53706, USA
}%

\author{Zhoudunming~Tu}
\email{zhoudunming@bnl.gov}
\affiliation{%
 Department of Physics, Brookhaven National Laboratory, Upton, New York 11973, USA
}%
\affiliation{Center for Frontiers in Nuclear Science, Stony Brook, New York 11794, USA}

\author{Raju~Venugopalan}
\email{rajuv@bnl.gov}
\affiliation{%
 Department of Physics, Brookhaven National Laboratory, Upton, New York 11973, USA
}%

\date{\today}
\begin{abstract}
Spin correlations of $\Lambda$-hyperons embedded in the QCD strings formed in high energy collider experiments 
provide unique insight into their 
locality and entanglement features.
We show from general considerations that while the Clauser-Horne-Shimony-Holt  inequality is less stringent for such states, 
they provide a  benchmark for quantum-to-classical transitions induced by varying  i) the associated hadron multiplicity, ii) the spin of nucleons, iii) the separation in rapidity between pairs, and iv) the kinematic regimes  accessed. These studies also enable the extraction of quantitative measures of quantum entanglement. We first explore such questions within a simple model of a QCD string composed of singlets of two partial distinguishable fermion flavors and compare analytical results to those obtained on quantum hardware. We further discuss a class of spin Hamiltonians that model the  
dynamics of $\Lambda$ spin correlations. 
Prospects for extracting  quantum features of QCD strings from hyperon measurements at current and future colliders are outlined. 

\end{abstract}


\maketitle

\newpage

The promise of solving {\it ab initio} real-time many-body problems in quantum field theory motivates the interest in quantum information science (QIS)
for high energy physics~\cite{Preskill:2018fag}. For instance, an outstanding problem in quantum chromodynamics (QCD) at high energies is the origin of 
``ridge"~\cite{Khachatryan:2010gv} long-range rapidity correlations which offer unique insight into the thermalization process in the quark-gluon plasma (QGP)~\cite{Berges:2020fwq}. While answers to these 
questions will only be obtained past the noisy intermediate scale quantum (NISQ) era, focused questions on 
problems universal to simpler systems can provide valuable answers sooner~\cite{Beane:2018oxh,Mueller:2019qqj,Lamm:2019uyc,Chakraborty:2020uhf,Kharzeev:2020kgc,Kreshchuk:2020dla,Liu:2020eoa,Briceno:2020rar,Davoudi:2020yln,deJong:2021wsd,Gustafson:2021imb,Barata:2020jtq,Barata:2021yri,Li:2021kcs,Honda:2021aum,Ikeda:2020agk,Klco:2021biu}. 

In the case of the ridge correlations, QIS studies may help identify (and classify) intrinsically quantum features such as Hanbury-Brown--Twiss and Bose-enhanced gluon correlations~\cite{Dumitru:2010iy,Kovner:2018azs}  arising from the entanglement of partons (quarks, antiquarks and gluons) within an ensemble of QCD strings 
~\cite{Kogut:1974ag,Bali:2000gf}. 
Their dynamics is presently only implemented classically in Monte Carlo (MC) generators that simulate  collider events~\cite{Andersson:1983ia}. Further, 
quantum correlations of partons are not easily separable from those arising from their rescattering~\cite{Dusling:2015gta} with increasing density in the string ensemble. Useful lessons on such  quantum-to-classical transitions may come from 
tabletop experiments with ultracold atomic gases~\cite{Joachim}; a powerful example of this synergy is nonthermal fixed points universal to QGP thermalization and ultracold atomic gases ~\cite{Berges:2014bba,Prufer:2018hto}.

Quantum correlations within QCD strings or string ensembles can be studied  at  electron-positron ($e^+ e^-$) colliders~\cite{Chen:2021eok}, in deeply inelastic electron-proton scattering (DIS) experiments~\cite{H1:2020zpd,Tu:2020pfl}, and in hadronic collisions~\cite{Tu:2019ouv}. For example, a remarkable observation in collider data that is suggestive of the role of entanglement is the apparent thermal distribution of  small numbers of produced particles in a QCD string~\cite{Berges:2017zws,Berges:2017hne}. 

In this paper, we will explore the possibility that $\Lambda$ and ${\bar \Lambda}$-hyperon spin correlations provide novel insight into intrinsically quantum features of many-body parton dynamics. Such measurements are feasible  because the weak decay  $\Lambda \to \pi^- + p$ allows one to extract the $\Lambda$'s (and analogously, that of ${\bar \Lambda}$ ) spin polarization to be  $\mathbf{P} = \alpha \mathbf{\hat a}$, where $\mathbf{\hat a}$ denotes the direction of the daughter proton's momentum in the $\Lambda$ rest frame, and 
 $\alpha \approx 0.750$~\cite{BESIII:2018cnd}\footnote{This value updates that of 0.642 quoted in Ref.~\cite{Commins:1983ns}.}  
 The use of $\Lambda$-hyperon spin correlations in QIS was first suggested in ~\cite{Tornqvist:1980af} as a test\footnote{It was further examined in the context of $e^+ e^-$ collisions \cite{HaoCPC2009} and for   top-quark and Higgs measurements at colliders~\cite{Barr:2021zcp,Fabbrichesi:2021npl}.} of 
 local hidden variable theory (LHVT)~\cite{EinsteinPR1935,Bohm:1951xw,Bohm:1951xx} by  employing the  Clauser-Horne-Shimony-Holt (CHSH)~\cite{PhysRevLett.23.880} inequality 
\begin{align} \label{eq:chsh-ineq}
    |E(\mathbf{\hat a}, \mathbf{\hat b}) - E(\mathbf{\hat a}, \mathbf{\hat b'})| + |E(\mathbf{\hat a'}, \mathbf{\hat b'}) + E(\mathbf{\hat a'}, \mathbf{\hat b})| \leq 2\,,
\end{align}
where 
$E(\mathbf{\hat a},\mathbf{\hat b}) = \langle \psi|\mathbf{\hat a} \cdot \mathbf{\sigma_1}\mathbf{\hat b} \cdot \mathbf{\sigma_2}|\psi\rangle$ 
with $\mathbf{\hat a}$ and $\mathbf{\hat b}$ corresponding to the momentum directions of the daughter particles\footnote{
Note that the Pauli operator $\mathbf{\sigma_i}$ acts on particle $i$, and we denote spin-up as 1 and spin-down as -1.}. Measured violations 
of the CHSH inequality imply a violation of LHVT. However, these Bell-type tests, while interesting, are unlikely to provide stronger constraints on LHVT beyond that of tabletop experiments\footnote{For discussions of precision LHVT tests, with emphases on potential loopholes, 
see~\cite{GiustinaPRL2015,HensenNature2015,RauchPRL2018}.}. Further, $\Lambda {\bar \Lambda}$ correlations in a QCD string reflect the dynamics of many-body mixed states for which the CHSH test in \cite{Tornqvist:1980af} is  inapplicable. Nevertheless, as we will discuss, a modified CHSH inequality, and related entanglement measures, can help quantify dynamical quantum-to-classical transitions in such systems. 

We will first demonstrate a general proof of a modified LHVT test for mixed states, discuss its relation to entanglement measures, and describe their possible implications for quantum-to-classical transitions. To address quantum dynamics in multiparticle production at colliders, we will first apply this framework to a simple string model and explore how hyperon correlations are washed away with increasing multiplicity. We then discuss a class of spin Hamiltonians that capture their underlying parton  dynamics. Simulations of  $\Lambda {\bar \Lambda}$ correlations are discussed next, with particular attention to their implementation  on quantum hardware. Finally, we address experimental opportunities in extracting quantum information from measurements. The Supplemental Material provides details of these measurements.


Consider the two-particle correlation function (or equivalently a joint probability distribution), 
\begin{align}\label{eq:corrfn}
    \frac{ \langle{n_\mathbf{\hat a}, n_\mathbf{\hat b}}\rangle}{\langle{n_\mathbf{\hat a}\rangle}{\langle n_\mathbf{\hat b}}\rangle} = \frac{P(\mathbf{\hat a}, \mathbf{\hat b})}{P(\mathbf{\hat a})P(\mathbf{\hat b})}\,,
\end{align}
where $n_\mathbf{\hat a}$ ($n_\mathbf{\hat b}$) indicates the number of daughter particles with momentum direction $\mathbf{\hat a}$ ($\mathbf{\hat b}$) measured in a single event, and $\langle\cdots\rangle$ denotes the ensemble average. The count pair in the numerator is from the same event; those in the denominator are from different events. 

We now discuss a theorem that encompasses how nonlocality and entanglement manifest in Eq.~(\ref{eq:corrfn}). For simplicity, we consider spins in the $x$-$z$ decay plane of the Bloch sphere and assume  $|\mathbf{P}| = 1$, indicating perfect discrimination between spin-up and spin-down. Thus a daughter proton (or pion) decaying along $\mathbf{\hat a}$ signifies a parent hyperon spin state of spin-up along $\mathbf{\hat a}$. The realistic determination of the spin direction is described later in the text and in the Supplemental Material.

The general spin state of the two spin-$\frac{1}{2}$ hyperons (with angles $\theta_a$ ($\theta_b$) of spin directions $\mathbf{\hat a}$ ($\mathbf{\hat b}$) relative to the z axis of the system can be represented by a density matrix $\rho_{ab}= \sum_{i = 1}^4 \sum_{j = 1}^4 \lambda_{ij} \ket{B_i}\bra{B_j}$  in the Bell basis, 
\begin{eqnarray}
\label{eq:bell}
     \ket{B_1} &=& \frac{\ket{00} + \ket{11}}{\sqrt{2}}\,;\,
    \ket{B_2} = \frac{\ket{00}-\ket{11}}{\sqrt{2}}\,; \nonumber \\
    \ket{B_3} &=& \frac{\ket{01} + \ket{10}}{\sqrt{2}}\,;\,
    \ket{B_4} = \frac{\ket{01} - \ket{10}}{\sqrt{2}}\,\,,
\end{eqnarray}
with $\lambda_{ij}$ real due to spins being in the $x$-$z$ plane.
Computing 
$P(\mathbf{\hat a}, \mathbf{\hat b})  = {\rm Tr}(\ket{\theta_a}\ket{\theta_b}\bra{\theta_a}\bra{\theta_b}\rho_{ab})$
and likewise, $P(\mathbf{\hat a}) = {\rm Tr}(\ket{\theta_a}\bra{\theta_a}\rho_a) $,
with $\rho_a = {\rm Tr}_b(\rho_{ab})$, and assuming both probabilities only depend on $\theta_a-\theta_b$ (rotational invariance, which sets $\lambda_{22}=\lambda_{33}$), and $P(\mathbf{\hat a}, \mathbf{\hat b})$ = $ P(\mathbf{\hat b}, \mathbf{\hat a})$ (which sets $\lambda_{ij}=0$, $i\neq j$),
we obtain 
\begin{align}\label{eq:cf-sym-rot}
    \frac{P(\mathbf{\hat a}, \mathbf{\hat b})}{P(\mathbf{\hat a})P(\mathbf{\hat b})} = 1 + (\lambda_{11} - \lambda_{44})\cos(\theta_a-\theta_b)
\end{align} 
We therefore conclude the following:\\
{\bf Theorem.} {\it A symmetric, rotationally invariant correlation function implies that the measured state $\rho_{ab}$ is diagonal in the Bell basis, with $\lambda_{22} = \lambda_{33}$.}

In the context of the generalized CHSH inequality~\cite{WuPRA2001,AndreevTMP2004}  for mixed two-particle spin-$\frac{1}{2}$ states diagonal  in the Bell basis (with $E(\mathbf{\hat a}, \mathbf{\hat b}) = E(\theta_a-\theta_b) \equiv E(\theta_{ab})$, and coplanar spin axes $\mathbf{\hat a}, \mathbf{\hat a'}, \mathbf{\hat b}, \mathbf{\hat b'}$),  Eq.~(\ref{eq:chsh-ineq}) reduces to the one-parameter inequality,
\begin{align}\label{eq:chsh-one-param-bell}
   \begin{cases}
    |E(\theta_{ab})| \leq  {\cal C}\,(1- \frac{2|\theta_{ab}|}{\pi})\,,\,\,\,\,\,\,\,\, |\theta_{ab}| \leq \frac{\pi}{2}\\
    |E(\theta_{ab})| \leq  {\cal C}\,(1 - \frac{2(\pi - |\theta_{ab}|)}{\pi})\,, \pi/2 \leq |\theta_{ab}| \leq \pi\,,
    \end{cases}
\end{align}
where ${\cal C} = |\lambda_{44}-\lambda_{11}|$. We can obtain $E(\theta_{ab})$ from the numerator of the measured two-particle correlation function as 
    $E(\theta_{ab}) = \frac{1}{4}(P(\mathbf{\hat a}, \mathbf{\hat b}) + P(\mathbf{-\hat a}, \mathbf{-\hat b}) - P(\mathbf{-\hat a}, \mathbf{\hat b}) -P(\mathbf{\hat a}, \mathbf{-\hat b})$, 
which gives, 
\begin{align}\label{eq:spincorr-sym-rot}
    E(\theta_{ab}) = (\lambda_{11} - \lambda_{44})\cos(\theta_{ab}) \,.
\end{align}
Comparing Eq.~(\ref{eq:spincorr-sym-rot}) to Eq.~(\ref{eq:chsh-one-param-bell}) 
leads us to a corollary to our theorem:\\
{\bf Corollary 1.} {\it A symmetric, rotationally invariant correlation function implies that the measured state $\rho_{ab}$ violates its related CHSH inequality, indicating incompatibility with LHVT.}

As the theorem and corollary indicate, the CHSH inequality is violated more easily for our mixed hyperon states relative to pure states. 
The above criteria are therefore sufficient to negate classical/deterministic explanations of their spin correlations.  

In quantifying entanglement, of several possible measures~\cite{Plenio:2007zz}, entanglement fidelity defined as 
\begin{align}\label{eq:ent-fid}
    \mathcal{F}_i = \bra{B_i}\rho_{ab}\ket{B_i}\equiv \lambda_{ii}\,,
\end{align}
is the most straightforward to extract from measured $\Lambda {\bar \Lambda}$ spin correlations with $\rho_{ab}$  entangled if $\mathcal{F}_i > \frac{1}{2}$. Since the coefficient of $\cos(\theta_{ab})$ in  Eq. \ref{eq:cf-sym-rot} is $\lambda_{11} - \lambda_{44}$, we obtain a second corollary to our theorem,\\
{\bf Corollary 2.} {\it If the magnitude of the coefficient of $\cos(\theta_{ab})$ in a symmetric rotationally invariant correlation function is $> \frac{1}{2}$, then the measured state $\rho_{ab}$ is entangled.}

The criterion $\mathcal{F}_i > \frac{1}{2}$ is  sufficient but not necessary for entanglement. A necessary and sufficient entanglement measure, albeit more challenging to extract at colliders, is the Peres-Horodecki positive partial transpose (PPT) 
criterion~\cite{Peres:1996dw,Horodecki:1997vt}; it is discussed at length in the Supplemental Material.

Our theorem and corollaries provide a novel approach to quantifying quantum-to-classical transitions in many-body systems. For example, the aforementioned ridge effect for  $\Lambda {\bar \Lambda}$ correlations could arise from  hydrodynamic flow (the LHVT) in high multiplicity events. 
As we would anticipate, rotational invariance of the correlations is broken because of a preferred ``reaction plane" in such events. The converse, a quantum effect that breaks rotational invariance is feasible; however since  microscopic 
interactions in QCD respect rotational invariance, its observed violation with increasing multiplicity signals onset of a quantum-to-classical transition.


To flesh out these general results, and in particular ascertain the role of entanglement, we will first consider a very simple spin model for $\Lambda$-hyperons embedded in a QCD string. Here heavy strange--antistrange quarks
($s {\bar s}$) are mixed up with light parton pairs of one  other flavor $u {\bar u}$ along the QCD string. Hadronization in this picture corresponds to the parton  ensemble of  $s, \bar s, u$ and $\bar u$  being grouped, after hadronization, into spin singlets with possible singlet combinations\footnote{These are proxies for $\Lambda {\bar \Lambda}$, kaon and pion states, respectively; up and down quark pairs are taken as indistinguishable.} being $s \bar s$,  $u \bar s$, $s \bar u$ and $u \bar u$. For $N$ partons,  there are $a$ singlets of type $s\bar s$, $b/2$ singlets of type $s \bar u$, $b/2$ singlets of type $u \bar s$, and $N/2 -a -b$ singlets of type $u \bar u$. Hence there are $2a + b$ particles of type $s$ or $\bar s$ and $N -2a -b$ particles of type $u$ or $\bar u$. 
We assume the ground state wave function of parton singlets corresponds to their occupying the lowest $N/2$ energy levels of the string, with $s\bar s$ on levels $1... a$, $s\bar u$ on levels $a+1... a + b/2$, $u \bar s$ on levels $a + b/2 + 1 ... a + b$, and $u \bar u$ on levels $a + b + 1... N/2$; the relevant aspect for us is the orthogonality of individual wave functions. 
\begin{figure}[h]
\centering
\includegraphics{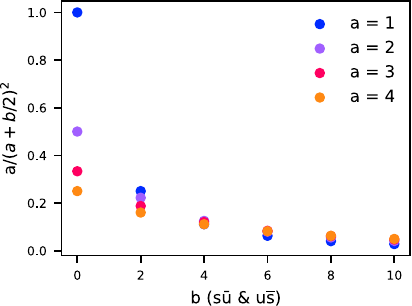}
      \caption{ The coefficient $\frac{a}{(a + b/2)^2}$ in Eq.~(\ref{eq:two-flavor}) plotted for various $a$ and $b$ corresponding to  $s\bar{s}$ and $u\bar{s}(s\bar{u})$ pairs in the string. A coefficient greater than $\frac{1}{2}$ satisfies the entanglement fidelity criterion.  
      }
      \label{fig:coeffss}
\end{figure}

 The explicit computation of $s \bar s$, $s{\bar u}$ and $u {\bar u}$ correlations is  worked out in the Supplemental Material; for the 
 $s{\bar s}$ ($\Lambda {\bar \Lambda}$) correlations, 
\begin{align}
\label{eq:two-flavor}
    \frac{P(\ket{\hat n_1}, \ket{\hat n_2})}{P(\ket{\hat n_1})P(\ket{\hat n_2})} = 1 - \frac{a}{(a + b/2)^2}\cos(\theta_2-\theta_1) \,.
\end{align}
This result depends on $a$ and $b$ since they are the only parameters that determine the number of $s \bar s$, $s \bar u$, and $u \bar s$ singlets. It  clearly violates the CHSH inequality for mixed states for all values of $a$ and $b$. The coefficient of the cosine is plotted for various $a$ and $b$ in Fig. \ref{fig:coeffss}.
As may be anticipated, adding mixed singlets, or having $b > 0$, decreases the entanglement fidelity since their indistinguishability washes out the spin correlation. The corresponding PPT entanglement criterion for this model is discussed in the Supplemental Material.

Our toy model shows that even though entanglement fidelity is washed away, nonlocality can persist; for locality to emerge, necessitates  further many-body interactions that break rotational invariance, generating classical correlations.


QIS discussions of stringy phenomena are dominantly in the Schwinger model and its variants~\cite{Berges:2017hne,Berges:2017zws}. However since  $\Lambda {\bar \Lambda}$ correlations are a promising probe of quantum features of strings, models where heavy strange quarks interact with light up/down quarks  provide novel insight. A good starting point is the Anderson model of localized impurities coupled to delocalized fermion spins~\cite{Keeling},
\begin{align}
    H_{\rm Anderson}&= \sum_{k\sigma}\epsilon_k a_{k\sigma}^\dagger a_{k\sigma} + \epsilon_d d_\sigma^\dagger d_\sigma + U d_{\uparrow}^\dagger d_{\uparrow} d_{\downarrow}^\dagger d_{\downarrow} \nonumber \\
    &+\frac{\eta}{\sqrt{V}}\left(d_\sigma^\dagger a_{k\sigma} + a_{k\sigma}^\dagger d_{\sigma} \right)\,.
 \end{align}
 Here the $d_\sigma^\dagger$($d_{\sigma}$)'s denote the localized impurities with spins $\sigma$ (denoting heavy strange quarks carrying $\Lambda$ spin~\cite{Burkardt:1993zh,Moretti:2019peg}) and the  $a_{\sigma}^\dagger$($a_\sigma$)'s represent the delocalized fermions (light up and down quarks); the first two terms in the Hamiltonian are their respective kinetic energies. The third term is a Hubbard-type hopping term for the light quarks and the final term denotes their spin coupling to the  strange quark impurities. This last term ``screens" the formation of $\Lambda {\bar \Lambda}$ singlets. A Schrieffer-Wolff transformation~\cite{Schrieffer-Wolff}, with the impurity kinetic energy $\epsilon_d$  below the Fermi energy recovers the Kondo Hamiltonian, describing the net spin coupling of delocalized fermions to the impurity spin ${\bf S}$. 
 
 Further insight into correlations between the $\Lambda {\bar \Lambda}$ pairs ``doping" the QCD string is obtained in an extension of the Kondo model\footnote{See \cite{Hattori:2015hka} for a similar discussion of heavy flavor impurities in quark matter at high baryon densities.}, whose ground state is a filled Fermi sea of light fermions and  impurities, with additional phase factors denoting  their spatial locations.  Eliminating excitations of the Fermi sea via another Schrieffer-Wolff transformation results in an effective Hamiltonian of localized impurities with interactions  mediated by the exchange of virtual electron-hole pairs. 
 
 This Ruderman-Kittel-Kasuya-Yoshida (RKKY) effective Hamiltonian~\cite{PhysRev.96.99,Kasuya,Yoshida}  mimics the QCD string at small values of the DIS Bjorken $x$ variable, where a large multiplicity of light quark/gluon pairs either screen or antiscreen the correlations between $\Lambda$-hyperons,  taking the form  
\begin{align}
 H_{\rm RKKY} = \sum_{jj^\prime} {\bf S}_j \cdot {\bf S}_{j^\prime}\,J_{\rm RKKY}(R_j - R_{j^\prime}) \,,
\end{align}
\noindent with
 $$J_{\rm RKKY}(R)=\frac{-J^2}{(k_F R)^4} [\sin{(2k_F R)} - 2k_F R \cos{(2k_F R)}]\,,$$
where $k_F$ is the Fermi momentum. It is ferromagnetic at short distances but has alternating sign at larger distances, suggestive of glassy dynamics. 
 \begin{figure}[h]
\centering
\includegraphics{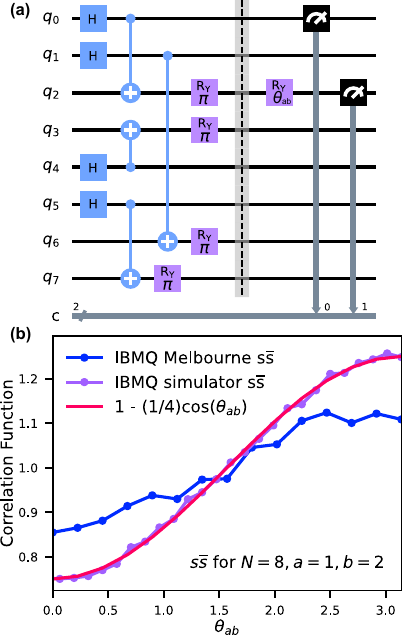} 
      \caption{(a) One of 16 circuits necessary to initialize and simulate the state of $N = 8$ particles with $a = 1$ $s\bar s$ singlets, $b = 2$ $s \bar u$ and $u \bar s$ singlets and one $u{\bar u}$ singlet. Here, each qubit $q_i$ carries the spin information of one of the $N = 8$ particles, while the classical register $c$ stores the value of the qubit obtained after measurement.  The barrier separates initialization from simulation of the correlation function. (b) Quantum simulation results for the $s {\bar s}$ correlation function compared to the prediction in Eq. \ref{eq:two-flavor}. Correlations obtained from the IBM Q Melbourne quantum computer are smaller than our analytical calculation due to quantum hardware noise; this hardware error will pose even greater restrictions for simulations of larger ensembles $N$.
      }
      \label{fig:qsim_812}
\end{figure}

While the RKKY model is a good model of $\Lambda{\bar \Lambda}$ correlations for small $x$, a better fit for the ``impurity doped" QCD string at large $x$ is the Anderson model with multiple impurities~\cite{_itko_2006}. In analogy to a quantum phase transition proposed~\cite{Eickhoff_2020} between Kondo and RKKY regimes, it would be interesting to investigate  consequences of the increased multiplicity of QCD strings with varying Bjorken $x$. In polarized DIS, valence quark spin plays an analogous role to a magnetic field providing an additional handle on simulating string dynamics. Thus mapping the rich dynamics of the Anderson/Kondo model, ``tuned" appropriately to measurements of $\Lambda {\bar \Lambda}$ correlations   embedded in QCD strings, offers a novel direction in QIS studies of hadronization at colliders.

There are several classical approaches to simulating the ground state properties of the aforementioned spin Hamiltonians~\cite{_itko_2006,Wilson1975,Wilson1980,PhysRevB.40.431,Anderson_1970}. However such Hamiltonians  suffer from a severe dynamical sign problem that afflicts the extraction of real-time correlations~\cite{PhysRevLett.115.266802}. Since the formation, evolution, and fragmentation of QCD strings are dynamical real-time problems, they are susceptible to the sign problem even in lower-dimensional incarnations. 

Quantum computers do not suffer from this problem, with benchmark computations performed for the Ising model in an external magnetic field \cite{Smith:2019mek}. The quantum computation of Anderson and Kondo lattices has been discussed previously~\cite{Garc_a_Ripoll_2008};  digital  simulations of these Hamiltonians, adapted to the QCD string, are in progress. 

As a first step, we wrote down quantum circuits for our toy model and performed  computations on IBM's QISKIT quantum simulator~\cite{Qiskit-Textbook} and on IBM Q quantum hardware, specifically 
\texttt{ibmq\_16\_melbourne}  containing 14 qubits~\cite{IBMQ}. This computation is outlined in the Supplemental Material. In Fig.~\ref{fig:qsim_812} (a), we show one of the 16 circuits necessary to generate and simulate the mixed spin density matrix for the case of $N=8$ spin-1/2 fermions, with $a=1$ (one $s{\bar s}$ pair), $b=2$ (one $s {\bar u}$ and ${\bar s} u$ singlet each) and $N/2-a-b$, one $u{\bar u}$ singlet.  In Fig.~\ref{fig:qsim_812} (b), we show the analytical result from Eq. \ref{eq:two-flavor} compared to the result from the QISKIT simulator; we find good agreement.  In contrast, the agreement with actual quantum hardware is not good, illustrating the challenge of reliable quantum simulation in the NISQ era. 

Finally, we discuss the experimental opportunities in measuring $\Lambda {\bar \Lambda}$ correlations at colliders. 
\begin{figure}[thb]
\centering
\includegraphics[width=0.9\linewidth]{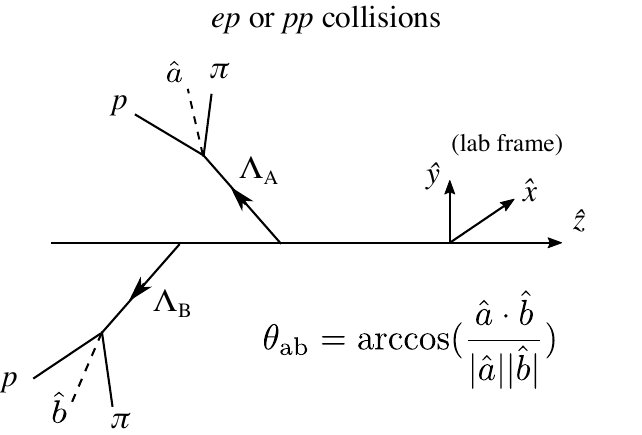}
  \caption{ Illustration of double $\Lambda$ polarization; here $\hat{a}$ ($\hat{b}$) denotes the momentum direction of  $\Lambda_A$ ($\Lambda_B$) daughter particle in the $\Lambda_A$ ($\Lambda_B$) rest frame.}
  \label{fig:measurements}
\end{figure}
 The $\Lambda$ and ${\bar \Lambda}$ spins are measured in terms of their  polarization, where the decay kinematics on an event-averaged basis reflects their spin projections~\cite{Guan:2018ckx,Bunce:1976yb,Heller:1978ty,Aleev:1982mz,Astier:2000ax,Astier:2001ve,Felix:2001zr,Adamovich:2004cc,Abt:2006da,SanchezLopez:2007aa,Agakishiev:2014kdy,Airapetian:2007mx,Hauenstein:2016som,STAR:2017ckg}. The CHSH inequality and entanglement measures are extracted from the correlation of their relative spin projections,
 illustrated in Fig.~\ref{fig:measurements}, and written as $N \propto 1+\alpha^{2} P_{\rm{\Lambda,\Lambda}}\cos{(n\theta_{\rm ab})}$, where $n$ is a free parameter that can be determined by the measurement and is expected to be less than unity due to a convolution between the intrinsic CHSH cosine modulation and the $\Lambda$ decay kinematics\footnote{The cosine modulation in Eq.~(\ref{eq:two-flavor}) and in $\Lambda$ decays are of different origin. }. As noted, 
$\alpha=0.750\pm0.010$~\cite{BESIII:2018cnd}, $\theta_{\rm ab}$ is the relative angle between daughter particles in their respective mother's rest frame and a nonzero $P_{\rm \Lambda,\Lambda}$ implies their spin correlation. 

Currently, no MC generators  implement spin entanglement at the parton level, providing a 
clear (null result) experimental baseline for entanglement searches. Specifically, we can simulate ``by hand" spin entanglement in the PYTHIA 8 MC event generator~\cite{Sjostrand:2007gs}; the Supplemental Material discusses in detail  simulation results and experimental measurements.

In summary, we derived in this paper
a modification  of the CHSH inequality, and related entanglement measures, for mixed states. These are  powerful tools in quantifying  quantum-to-classical transitions in the many-body dynamics of strings in the collider environment. We further constructed theoretical models to capture the quantum dynamics of QCD strings with embedded hyperons and discussed how  these can be extracted from $\Lambda {\bar \Lambda}$ correlations. 
With a longer term view of QIS, we performed first simulations on quantum hardware; these provide a benchmark and illustrate the current challenges in reliable extraction of quantum information. Further systematic studies implementing quantum error correction,  state preparation, Trotter evolution, and entanglement measures,  will be reported separately. MC simulations of  $\Lambda$ correlation measurements at colliders suggest that prospects for extracting information on quantum-to-classical transitions in QCD strings are promising.

\begin{acknowledgments}
We thank Robert Konik for discussions on spin Hamiltonians, 
Elke-Caroline Aschenauer for discussions on spin polarization measurements and Alexander Jentsch for discussions of $\Lambda$ particle detection in the far-forward region at the EIC. We acknowledge the use of IBM Quantum services for this work. The views expressed are those of the authors, and do not reflect the official policy or position of IBM or the IBM Quantum team.

The work of W.G. is supported in part by the U.S. Department of Energy, Office of Science, Office of Workforce Development for Teachers and Scientists (WDTS) under the Science Undergraduate Laboratory Internships Program (SULI). Access to IBM's quantum hardware is provided through Harvard University and Brookhaven National Laboratory's Scientific Data and Computing Center (SDCC).   

The work of G.P. was supported in part by the S.N. Bose Scholars Program, jointly funded by the Government of India [Department of Science and Technology (DST) - Science and Engineering Board (SERB)], Indo-U.S. Science and Technology Forum (IUSSTF) and WINStep Forward and in part by the U.S. Department of Energy under Award No.~DE-SC0012704.

The work of Z.T. is supported by LDRD-039 and the Goldhaber Distinguished Fellowship at Brookhaven National Laboratory. Z.T. and R.V. are supported by the U.S. Department of Energy under Award No.~DE-SC0012704. 

The work of W.G. and R.V. on quantum information science is supported in part by the U.S. Department of Energy, Office of Science National Quantum Information Science Research Center's Co-design Center for Quantum Advantage (${\rm C}^2$QA) under Award No.~DE-SC0012704.

\end{acknowledgments}

\appendix

\section{PPT measure for QCD strings}\label{sec:Appendix A}

For mixed states, the Peres-Horodecki criterion~\cite{Peres:1996dw,Horodecki:1997vt}, also known as the positive partial transpose (PPT) criterion, can be used to detect separability. This criterion is a necessary condition\footnote{It is also sufficient if the dimension of the product space is  $2 \times 2$ or $2 \times 3$.} for the separability of the joint density matrix $\rho$ of two systems $A$ and $B$. We first describe this criterion and then briefly discuss observables that can signal its violation, indicating bipartite entanglement between $A$ and $B$.

If we have $\rho$ on $\mathcal{H}_A \otimes \mathcal{H_B}$,
\begin{align}
    \rho = \sum_{ijkl} p_{kl}^{ij} \ket{i}\bra{j} \otimes \ket{k}\bra{l}\,,
\end{align}
where $\ket{i}, \ket{j}$ ($\ket{k}, \ket{l}$) label an orthonormal basis for $A$ ($B$), the partial transpose is defined as follows:
\begin{align}
    \rho^{T_B} &= (I\otimes T)\rho \notag \\
    &= \sum_{ijkl}p_{kl}^{ij} \ket{i}\bra{j} \otimes (\ket{k}\bra{l})^T \notag \\
    &=  \sum_{ijkl}p_{kl}^{ij} \ket{i}\bra{j} \otimes \ket{l}\bra{k}\,.
\end{align}
If $\rho$ is separable, then all the eigenvalues of $\rho^{T_B}$ are positive. If $\rho^{T_B}$ has one or more negative eigenvalues, then $\rho$ is entangled.

It is enlightening to first consider this criterion on a Werner state~\cite{PhysRevA.40.4277}, or a two-particle mixed state consisting of maximally entangled and maximally mixed components:
\begin{align}\label{eq:2-werner}
    \rho_{AB} &= p \ket{S_{AB}}\bra{S_{AB}} + \frac{(1-p)}{4}I_A \otimes I_B \,,\\
    \rho_{A} &= \frac{1}{2}I_A\,.
\end{align}
We note that our two-particle reduced density matrix in the two-flavor case, Eq. \ref{eq:rho2-2flav}, adopts precisely this form. Then, taking the partial transpose over subspace $B$, we find that the eigenvalues of $\rho_{AB}^{T_B}$ are
\begin{align}
    \nu_1 = \frac{1 + p}{4}\,\,;\,\,
    \nu_2 = \frac{1 - 3p}{4}\,.
\end{align}
Here $\nu_1$ has multiplicity 3. Since a negative eigenvalue indicates entanglement, we deduce that $\rho$ is entangled for $1/3 <p \leq 1$.

As the two-flavor two-particle reduced density matrix is of the specific form of a Werner state, we see that the condition for entanglement according to PPT here-- that the coefficient of the cosine in Eq.~8 of the main manuscript be greater than $\frac{1}{3}$-- is actually less restrictive than that predicted by our general theorem, which requires the coefficient to be greater than $\frac{1}{2}$. 

We can now question whether accordance with the PPT criterion can be detected via the measurement of observables in a high energy collision. It is important to note here that if $\rho^{T_B}$ does indeed have negative eigenvalues, then  $\rho^{T_B}$ is not a physical state, and thus the act of partial transposition does not correspond to a physical process. However, for any observable $\hat A$ \cite{Band2018},
\begin{align}
    \braket{\hat A}_{\rho^{T_B}} = \braket{\hat A^{T_B}}_\rho\,.
\end{align} 
Thus if we can find a positive operator $\hat A^2$, with corresponding $(\hat A^2)^{T_B}$,  such that
\begin{align}
    \braket{\hat A^2}_{\rho^{T_B}} = \braket{(\hat A^2)^{T_B}}_\rho < 0\,,
\end{align}
 this then indicates that $\rho^{T_B}$ is not positive; hence $\rho$ is entangled by the PPT criterion.
For two-particle density matrices of the form of the Werner state in Eq. \ref{eq:2-werner}, we consider
\begin{align}\label{eq:opA}
     \hat A &=  I_1 \otimes  I_2 +\sigma_{1 x} \sigma_{2 x} - \sigma_{1 y} \sigma_{2 y} + \sigma_{1 z}\sigma_{2 z} \,,\\
     \hat A^2 &= 4 I_1 \otimes I_2 
     - 4(-\sigma_{1 x} \sigma_{2 x} +\sigma_{1 y} \sigma_{2 y} - \sigma_{1 z}\sigma_{2 z})\,, \\
     (\hat A^2)^{T_2} &= 4 I_1 \otimes I_2 - 4(-\sigma_{1 x} \sigma_{2 x} -\sigma_{1 y} \sigma_{2 y} - \sigma_{1 z}\sigma_{2 z})\,.
 \end{align}
For the $\rho$ given in Eq. \ref{eq:2-werner}, 
\begin{align}
    \braket{\hat A^2}_{\rho^{T_B}} &= \braket{(\hat A^2)^{T_B}}_{\rho} = \Tr((\hat A^2)^{T_2}\rho) \notag \\ &= 4 - 4\Tr(3p\ket{S_{AB}}\bra{S_{AB}}) 
    = 4-12p \,.
\end{align}
This becomes negative for $p > \frac{1}{3}$, as we expect from our previous work. Again, as the reduced two-particle density matrix in the two-flavor case is a Werner state, using $\hat A$ given in Eq. \ref{eq:opA} to assess entanglement via the PPT criterion would also yield that the state is entangled when the coefficient of the cosine in Eq.~8 of the main manuscript is greater than $\frac{1}{3}$.

\section{Details of two flavor spin chain model}\label{sec:Appendix B}

We provide here details of the computation of the two flavor spin chain model containing four different types of particles: $s, \bar s, u$ and $\bar u$ comprising $N$ total particles, where there are $a$ singlets of type $s\bar s$, $b$ singlets of type $s \bar u$ or $u \bar s$, and $N/2 -a -b$ singlets of type $u \bar u$. 


We can construct an appropriate wavefunction (antisymmetric under exchange of $s$ with $s$, $ \bar s$ with $\bar s$, $u$ with $u$, and $\bar u$ with $\bar u$) in which $s\bar s$, $s \bar u$, $u \bar s$, and $u \bar u$ are paired into singlets:
\begin{align}
    &\ket{\Psi} \sim  \sum_{\sigma(s)}\sum_{C(\bar s)}\sum_{\sigma(u)}\sum_{C(\bar u)}
    \sgn(\sigma(s))\sgn(\sigma(u)) \notag \\
    &\times \underbrace{\ket{S_{\sigma(1), C(a + \frac{b}{2} + 1)}...}}_{a \;\; s\bar s} \underbrace{\ket{S_{\sigma(a+1), C(a + \frac{b}{2} + \frac{N}{2} + 1)}...}}_{\frac{b}{2} \;\; s\bar u} \notag \\
    &\times \underbrace{\ket{S_{C(2a + \frac{b}{2} + 1), \sigma(2a + b + 1)}...}}_{\frac{b}{2} \;\; \bar s u}\underbrace{\ket{S_{\sigma(2a + \frac{3b}{2} + 1), C(a + b + \frac{N}{2} + 1)}...}}_{\frac{N}{2} - a - b \;\; u\bar u} \notag \\
    &\times\Phi^{+}(X) \,,
\end{align}
\noindent with $X=(\sigma(1), \sigma(a + \frac{b}{2} + 1), ..., (\sigma(2a + \frac{3b}{2} + 1), \sigma(a + b + \frac{N}{2} + 1))$. The expression above denotes the sum over permutations of the $s$ and $u$ particles, as well as the sum over combinations of the $\bar s$ and $\bar u$ particles. For example, we freely permute particles $1 ... a + b/2$, or the $s$ particles. However, for the $a + b/2$ total $\bar s$ particles, we must choose $\begin{pmatrix} a + b/2 \\ a \end{pmatrix}$ $a$ particles to pair with the permuted $s$ particles, leaving the rest to pair with the permuted $u$ particles. In this manner, we count all the possible permutations that account for indistinguishable $s$,  $\bar s$, $u$, and $\bar u$ while keeping the required singlet structures. 

The $\Phi^+$ operator completely symmetrizes the spatial wavefunctions across all the pairs $(\sigma(1), \sigma(a + \frac{b}{2} + 1))...(\sigma(2a + \frac{3b}{2} + 1), \sigma(a + b + \frac{N}{2} + 1))...$ as follows:
\begin{align}
   &\Phi^+((1, \sigma(a + \frac{b}{2} + 1))...) \sim\nonumber\\
   &...\phi^\alpha(\mathbf{x_{\sigma(1)}})\phi^\alpha(\mathbf{x_{\sigma(a + \frac{b}{2} + 1)}})
   \phi^\beta(\mathbf{x_{\sigma(2)}})\phi^\beta(\mathbf{x_{\sigma(a + \frac{b}{2} + 2)}})... \notag \\
    &+ ...\phi^\alpha(\mathbf{x_{\sigma(2)}})\phi^\alpha(\mathbf{x_{\sigma(a + \frac{b}{2} + 2)}})
    \phi^\beta(\mathbf{x_{\sigma(1)}})\phi^\beta(\mathbf{x_{\sigma(a + \frac{b}{2} + 1)}})...\,,
\end{align}
where the superscripts of the single particle spatial wavefunctions, $\{\phi^\alpha \;|\; \alpha \in \{1, 2, ..., \frac{N}{2}\} \}$,  
denote the different pair energy levels. Thus the pairs  $(\sigma(1), \sigma(a + \frac{b}{2} + 1))...(\sigma(2a + \frac{3b}{2} + 1), \sigma(a + b + \frac{N}{2} + 1))...$ all each have an equal probability of occupying any level $1$ through $\frac{N}{2}$. 

The full density matrix of the system is  given by $\rho = \ket{\Psi}\bra{\Psi}$.  However since in our simple model we only care about the spin subsystem,  we compute the reduced density matrix $\rho_{\rm spin}$ by tracing out the spatial degrees of freedom:
$\rho_{\rm spin} = {\rm Tr}_{\rm spatial}(\ket{\Psi}\bra{\Psi})$.

 If we assume that the spatial wavefunctions are orthogonal,  $\braket{\phi^\alpha(\mathbf{x_{i})} |\phi^{\beta}(\mathbf{x_{i})}}$ = 0 for $\alpha \neq \beta$, the symmetrized spatial wavefunctions are then also orthogonal. This allows us to simplify the spin density matrix to 
 \begin{align}\label{eq:2flav-spin}
    &\rho_{\rm spin} = \frac{1}{{\cal N}} \sum_{\sigma(s)}\sum_{C(\bar s)}\sum_{\sigma(u)}\sum_{C(\bar u)}\nonumber \\
    &\times \underbrace{\ket{S_{\sigma(1), C(a + \frac{b}{2} + 1)}...}}_{a \;\; s\bar s}  \underbrace{\bra{S_{\sigma(1), C(a + \frac{b}{2} + 1)}...}}_{a \;\; s\bar s} \notag \\ &\times \underbrace{\ket{S_{\sigma(a+1), C(a + \frac{b}{2} + \frac{N}{2} + 1)}...}}_{\frac{b}{2} \;\; s\bar u} \underbrace{\bra{S_{\sigma(a+1), C(a + \frac{b}{2} + \frac{N}{2} + 1)}...}}_{\frac{b}{2} \;\; s\bar u} \notag \\
    &\times \underbrace{\ket{S_{C(2a + \frac{b}{2} + 1), \sigma(2a + b + 1)}...}}_{\frac{b}{2} \;\; \bar s u}\underbrace{\bra{S_{C(2a + \frac{b}{2} + 1), \sigma(2a + b + 1)}...}}_{\frac{b}{2} \;\; \bar s u} \notag \\
    &\times\underbrace{\ket{S_{\sigma(2a + \frac{3b}{2} + 1), C(a + b + \frac{N}{2} + 1)}...}}_{\frac{N}{2} - a - b \;\; u\bar u} \underbrace{\bra{S_{\sigma(2a + \frac{3b}{2} + 1), C(a + b + \frac{N}{2} + 1)}...}}_{\frac{N}{2} - a - b \;\; u\bar u}
\end{align}
where the normalization 
\begin{align}
    {\cal N} = &\left(a + \frac{b}{2}\right)! \left(\frac{N}{2} - a - \frac{b}{2}\right)! 
\begin{pmatrix}a + \frac{b}{2} \\ a\end{pmatrix} \begin{pmatrix}\frac{N}{2} -a - \frac{b}{2} \\ \frac{b}{2} \end{pmatrix}\,. \notag \\
\end{align}

To compute $s {\bar s}$ correlations, we must first compute the reduced density matrix:
$\rho_{1, a + \frac{b}{2} + 1} = {\rm Tr}_{\text{rest}}(\rho_{\rm spin})$. To do so, we consider terms of type $\ket{S_{1, a + \frac{b}{2} + 1}}\bra{S_{1, a + \frac{b}{2} + 1}}$ and 
$\ket{S_{1l}}\ket{S_{a + \frac{b}{2} + 1, k}}\bra{S_{1l}}\bra{S_{a + \frac{b}{2} + 1,k}}$. For $\ket{S_{1, a + \frac{b}{2} + 1}}\bra{S_{1, a + \frac{b}{2} + 1}}$, the contribution to the reduced density matrix
is
\begin{align}
    \frac{a}{(a + \frac{b}{2})^2} \ket{S_{1, a + \frac{b}{2} + 1}}\bra{S_{1, a + \frac{b}{2} + 1}}\,.
\end{align}
Likewise, for $\ket{S_{1l}}\ket{S_{a + \frac{b}{2} + 1, k}}\bra{S_{1l}}\bra{S_{a + \frac{b}{2} + 1,k}}$, the contribution to the reduced density matrix is 
\begin{align}
\frac{(a + \frac{b}{2})^2 - a}{(a + \frac{b}{2})^2}\, \frac{1}{4}\, I_1 \otimes I_{a + \frac{b}{2} + 1}\,.
\end{align}
Thus the net contribution to the reduced density matrix is
\begin{align}\label{eq:rho2-2flav}
    \rho_{1, a + \frac{b}{2} + 1} &= \frac{a}{(a + \frac{b}{2})^2} \ket{S_{1, a + \frac{b}{2} + 1}}\bra{S_{1, a + \frac{b}{2} + 1}} \nonumber \\ &+ \frac{(a + \frac{b}{2})^2 - a}{(a + \frac{b}{2})^2} \frac{1}{4} I_1 \otimes I_{a + \frac{b}{2} + 1}\,.
\end{align}
With this, we can calculate the two particle probability,
\begin{align}
    &P(\ket{\hat n_1}, \ket{\hat n_2}) = \Tr(\ket{\theta_1\theta_2}\bra{\theta_1\theta_2}\rho_{1, a + \frac{b}{2} + 1})\notag\\
    &=  \frac{a}{2(a + \frac{b}{2})^2}\sin(\frac{\theta_2-\theta_1}{2})^2 + \frac{(a + \frac{b}{2})^2 - a}{4(a + \frac{b}{2})^2}\,,
\end{align}
where $\ket{\hat n} = \ket{\theta}=  \cos(\theta/2) \ket{0} + \sin(\theta/2) \ket{1}$.

Similarly, the single particle reduced density matrix is
\begin{align}
    {\rm Tr}_{a + \frac{b}{2} + 1}(\rho_{1, a + \frac{b}{2} + 1}) &= \frac{a}{2(a + \frac{b}{2})^2} I_1 + \frac{(a + \frac{b}{2})^2 - a}{2(a + \frac{b}{2})^2} I_1 \nonumber \\
    &= \frac{1}{2}I_1\,,
\end{align}
giving $P(\ket{\hat n_1}) = \frac{1}{2}$. Thus the correlation function for $s\bar s$ is
\begin{align}
    \frac{P(\ket{\hat n_1}, \ket{\hat n_2})}{P(\ket{\hat n_1})P(\ket{\hat n_2})} = 1 - \frac{a}{(a + b/2)^2}\cos(\theta_2-\theta_1)\,,
\end{align}
which is the result reported in Eq.~8.
This result only depends on $a$ and $b$, which makes sense, since $a$ and $b$ are the only parameters that determine the number of $s \bar s$, $s \bar u$, and $u \bar s$ singlets. 

Using the same procedure, we also calculate the  $s \bar u (u \bar s)$ correlations:
\begin{align}
    \frac{P(\ket{\hat n_1}, \ket{\hat n_2})}{P(\ket{\hat n_1})P(\ket{\hat n_2})} = 1 - \frac{b/2}{(a + \frac{b}{2})(\frac{N}{2}-a -\frac{b}{2})}\cos(\theta_2-\theta_1) \,,
\end{align}
and likewise for $u \bar u$,
\begin{align}
\frac{P(\ket{\hat n_1}, \ket{\hat n_2})}{P(\ket{\hat n_1})P(\ket{\hat n_2})} = 1 - \frac{(N/2-a -b)}{(\frac{N}{2}-a -\frac{b}{2})^2}\cos(\theta_2-\theta_1)\,.
\end{align}
    
\section{Implementing the two flavor spin chain model on quantum hardware}\label{sec:Appendix C}

To simulate mixed states $\rho$ on quantum hardware, 
\begin{align}\label{eq:mixed-state}
    \rho = \sum_{i} p_i \ket{\psi_i}\bra{\psi_i}\,,
\end{align}
we adopt the straightforward approach of initializing and testing each constituent pure state $\ket{\psi_i}\bra{\psi_i}$, and then weighting the results by their appropriate probabilities $p_i$. 

As a representative case of the two flavor spin chain model, we specifically consider an ensemble of $N = 8$ particles, with $a = 1$ $s\bar s$ singlets, $b = 2$ total $s\bar u$ and $u \bar s$ singlets, and one $u \bar u$ singlet. From Eq. \ref{eq:2flav-spin}, we see that our initial state is then given by the spin density matrix:
\begin{align}\label{eq:rho-812}
    \rho = \frac{1}{16}( &\ket{S_{13}S_{27}S_{45}S_{68}}\bra{S_{13}S_{27}S_{45}S_{68}} \notag \\ + & \ket{S_{13}S_{27}S_{46}S_{58}}\bra{S_{13}S_{27}S_{46}S_{58}} \notag \\ + &\ket{S_{23}S_{17}S_{45}S_{68}}\bra{S_{23}S_{17}S_{45}S_{68}} \notag \\ + & \ket{S_{23}S_{17}S_{46}S_{58}}\bra{S_{23}S_{17}S_{46}S_{58}}  \notag \\
    + &\ket{S_{14}S_{27}S_{35}S_{68}}\bra{S_{14}S_{27}S_{35}S_{68}}\notag \\ + & \ket{S_{14}S_{27}S_{36}S_{58}}\bra{S_{14}S_{27}S_{36}S_{58}}  \notag \\ + &\ket{S_{24}S_{17}S_{35}S_{68}}\bra{S_{24}S_{17}S_{35}S_{68}} \notag \\ + &\ket{S_{24}S_{17}S_{36}S_{58}}\bra{S_{24}S_{17}S_{36}S_{58}} \notag \\
    +&\ket{S_{13}S_{28}S_{45}S_{67}}\bra{S_{13}S_{28}S_{45}S_{67}} \notag \\ + &\ket{S_{13}S_{28}S_{46}S_{57}}\bra{S_{13}S_{28}S_{46}S_{57}}   \notag \\ + &\ket{S_{23}S_{18}S_{45}S_{67}}\bra{S_{23}S_{18}S_{45}S_{67}} \notag \\ + &\ket{S_{23}S_{18}S_{46}S_{57}}\bra{S_{23}S_{18}S_{46}S_{57}} \notag \\
    + &\ket{S_{14}S_{28}S_{35}S_{67}}\bra{S_{14}S_{28}S_{35}S_{67}}  \notag \\ + & \ket{S_{14}S_{28}S_{36}S_{57}}\bra{S_{14}S_{28}S_{36}S_{57}} \notag \\ + &\ket{S_{24}S_{18}S_{35}S_{67}}\bra{S_{24}S_{18}S_{35}S_{67}} \notag \\ + & \ket{S_{24}S_{18}S_{36}S_{57}}\bra{S_{24}S_{18}S_{36}S_{57}})\,.
\end{align}
The initialization of the first term in Eq. \ref{eq:rho-812} on a quantum circuit is shown in Fig.~2~(a) of the main manuscript. 

To simulate a correlation function measurement of Eq.~8 of the main manuscript-- the second part of the circuit in Fig.~2~(a) of the main manuscript-- we first select two representative qubits to probe. For $s \bar s$ correlations, we choose qubits $q_0$, representing a $s$ spin, and $q_2$, representing a $\bar s$ spin. Note that as the spins are indistinguishable within the same particle type, our particular selection of qubits does not matter. We then rotate one of the selected qubits, $q_2$ in our case, by a varying $\theta_{ab}$ about the $y$ axis, thus demonstrating varying spin polarizations projected into the $x-z$ plane. 

For each $\theta_{ab}$, we measure $q_0$ and $q_2$ in the $z$ basis for a total of 1000 shots. The joint probability that both qubits point in the same direction is thus $n_{11}/1000$, while the individual probabilities are $n_{10}/1000$ and $n_{01}/1000$, where $n_{11}$ indicates the counts of the measurement result $'11'$. For each $\theta_{ab}$, we thus see from the general form of the correlation function Eq.~2 of the main manuscript that the correlation is
\begin{align}
    1000\cdot\frac{n_{11}}{n_{10} n_{01}}\,.
\end{align}
Averaging this across all $16$ circuits, each representing a term in Eq. \ref{eq:rho-812}, for each $\theta_{ab}$ value produces the plot in Fig.~2~(b). 

\begin{figure*}
    \centering
    \includegraphics[width=0.6\linewidth]{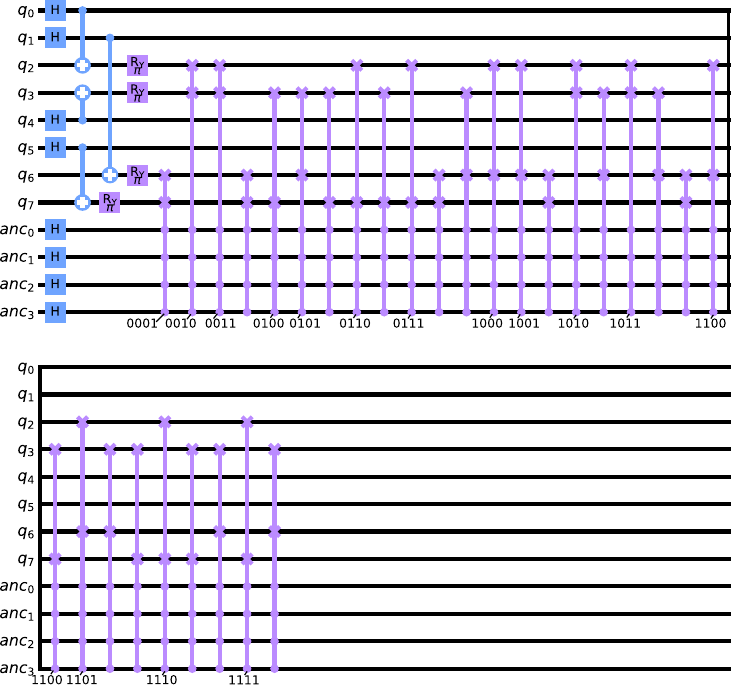}
    \caption{A circuit that initializes Eq. \ref{eq:rho-812} in the main quantum register. Swaps controlled on each of $i \in [0, 16]$, where $i$ counts in binary, are applied to initialize a full state of the form Eq. \ref{eq:mixed-state-full}. The binary form of $i$ for each group of cSWAPs is listed in the circuit.}
    \label{fig:cfull-812}
\end{figure*}

Though this method of simulating mixed states via constructing pure states and averaging is straightforward, it can be cumbersome and redundant as the same simulation must be run multiple times for a single result. Instead, it is useful to note a mixed state Eq. \ref{eq:mixed-state} can also be initialized by constructing the pure state
\begin{align}\label{eq:mixed-state-full}
    \rho' = \sum_{i = 0}^{\mathcal{N}-1} \sqrt{p_i} \ket{\psi_i} \ket{i}\,.
\end{align}
Here $\mathcal{N}$ is the total number of terms in the mixed state, and $i$ counts over the terms in binary. After initializing $\rho'$, if we then only work in the subspace spanned by the $\ket{\psi_i}$, we effectively have the mixed state Eq. \ref{eq:mixed-state}. As an example, we can construct our ensemble $N = 8$, $a = 1$, $b = 2$ using this method. The requisite circuit is given in Fig. \ref{fig:cfull-812}. For our states, we need a total of $\log_2(\mathcal{N})$ ancilla initialized in an equal superposition of $\ket{0}$ to $\ket{\mathcal{N}}$. We then perform cSWAP controlled on each of $i$ on the ancilla, $i \in [0, \mathcal{N}]$, targeting the necessary qubits needed to swap from one configuration of permutations to another. Note that though this way of initialization is more concise, it has a much higher cost of multi-qubit gates.

\section{Measuring $\Lambda {\bar \Lambda}$ correlations at collider energies\label{sec:Appendix D}}

The experimental method of measuring $\Lambda {\bar \Lambda}$ correlation in high energy collisions is similar across all available systems ($e^{+}e^{-}$, $ep$ DIS, and $pp$ collisions). Therefore we will  use $ep$ DIS to elaborate on the experimental techniques and detector requirements. 

\begin{figure}[h]
\centering
\includegraphics[width=0.79\linewidth]{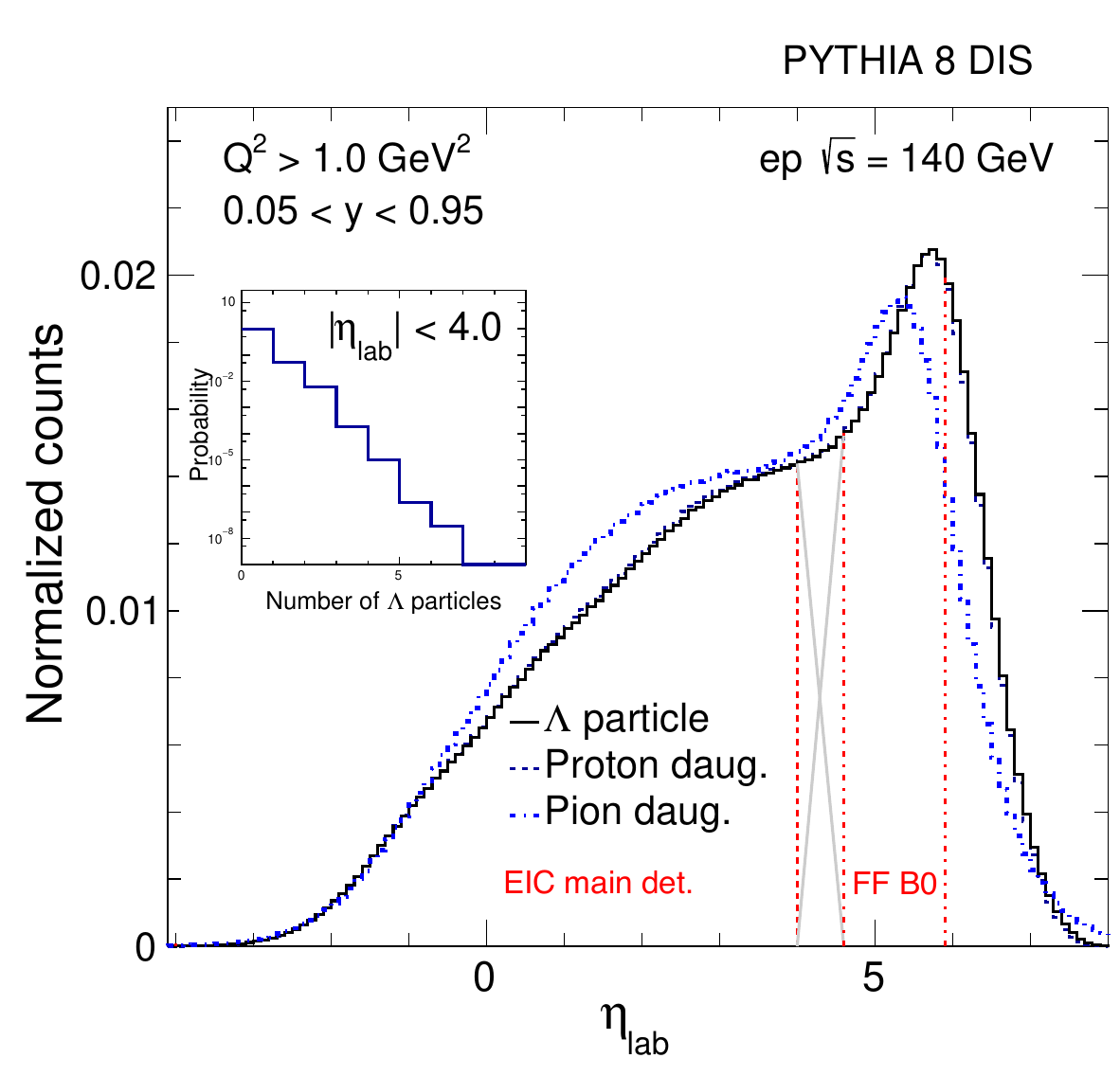}
      \caption{Pseudorapidity distributions of $\Lambda$ particles and their decays in $ep$ DIS events using the PYTHIA 8 model~\cite{Sjostrand:2007gs}. The acceptance of the EIC main detector and the Far-forward region B0 tracker are shown by the red dashed line. The number of $\Lambda$ particles per event is also shown.
      }
      \label{fig:figure_paper_2}
\end{figure}

Fig.~\ref{fig:figure_paper_2} shows two $\Lambda$ (either $\Lambda$ or $\bar \Lambda$) particles produced in a high energy $ep$ DIS collision decay into protons and pions. The daughter momentum can be reconstructed using particle detectors and boosted back to the rest frame of the $\Lambda$ particle, where either daughter particle can be used. In the rest frame of $\Lambda$, the momenta vector of the daughter are represented by $\hat{a}$ and $\hat{b}$ for $\Lambda_{\rm A}$ and $\Lambda_{\rm B}$, respectively. The nature of the weak decay of $\Lambda$ particle will reflect the $\Lambda$ spin in terms of the decay angular distribution in the rest frame. In the simplest (albeit unrealistic) scenario, the $\hat{a}$ and $\hat{b}$ are the spin directions of $\Lambda_{\rm A}$ and $\Lambda_{\rm B}$, which is considered in the model discussed earlier in this Letter. However, in reality, the direction of the decay is smeared by an angular distribution as 
\be
\frac{dN}{d\cos{(\theta^{\ast}})}\propto 1+\alpha P_{\rm{\Lambda}}\cos{(\theta^{\ast}})\,.
\ee
\noindent Here the hyperon decay parameter $\alpha=0.750\pm0.010$~\cite{BESIII:2018cnd}, and the angle $\theta^{\ast}$ is the angle between the decay daughter and the spin projection axis in the rest frame of $\Lambda$. 

There are two key aspects to consider in the experimental measurement. First, the spin of the $\Lambda$ is measured through an angular distribution, and thus not event-by-event accessible. Even if the weak constant is unity and the spin is perfectly aligned with the projected axis, the decay angular distribution is still a cosine modulation with its amplitude at the maximum. It is important to note that this cosine modulation is different from that in Eq.~8 of the main manuscript, which is a property of CHSH inequality. Therefore, the underlying correspondence of the two could be directly linked and  constrained by the experimental measurement, e.g., the determination of the parameter $n$ mentioned in the main text. 

Secondly, the weak decay constant sets the maximum measurable modulation of the cosine distribution, and in particular $\alpha^{2}\sim50\%$ in $\Lambda {\bar \Lambda}$ correlations because there are two decays involved. Therefore in order to properly compare the experimental data with our model or quantum simulations on a digital quantum computer, the nature of the $\Lambda$'s weak decay will have to be taken into account. Since the measurable signal is weaken by the weak decay, the requirement of statistics to extract the polarization or entanglement fidelity becomes higher to obtain the same statistical significance comparing to the ideal case ($\alpha = 1$).

In order to demonstrate the potential experimental opportunity of the double $\Lambda$ polarization, the first step is to check the experimental method in MC generators. Here we expect no classical mechanism can result in the violation of the modified CHSH inequality in this study. We use the PYTHIA 8 Monte Carlo (MC) event generator to simulate the $ep$ DIS events with $\Lambda$ particles for kinematic phase space $1<Q^{2}<100~\rm{GeV^{2}}$ and $0.05<y<0.95$. Here $Q^{2}$ is the virtuality of the emitted virtual photon and $y$ is the inelasticity of the DIS event. A total of 4 billion events are generated, corresponding to an integrated luminosity of $\sim3~\rm{fb^{-1}}$. The energy configuration is chosen at the top energy for $ep$ collisions at the Electron-Ion Collider~\cite{Accardi:2012qut}--18 GeV electrons scattering off 275 GeV proton beams. In this study, the detector and beam related effects, for instance the inefficiency of the detector or the angular divergence of the beam, are not implemented. However a basic detector acceptance is used based on EIC design~\cite{AbdulKhalek:2021gbh},  $|\eta_{\rm{lab}}|<4.0$ and $p_{\rm T}>150~\rm MeV$, for accepting the $\Lambda$ particles and their decays in the main detector. In Fig.~\ref{fig:figure_paper_2}, the pseudorapidity distributions of $\Lambda$ particles and their decay products in $ep$ DIS events for center-of-mass energy 140 GeV are shown. The acceptance of the main detector and in the far-forward (FF) region using the B0 tracker (FF B0) is indicated by the red dashed line. In the inset panel, the number of $\Lambda$ particles is shown per event within the EIC acceptance. Similar measurements can be performed using  $pp$ collisions at Relativistic Heavy-Ion Collider (RHIC) and at the Large Hadron Collider.

\begin{figure}[tbh]
\centering
\includegraphics[width=0.8\linewidth]{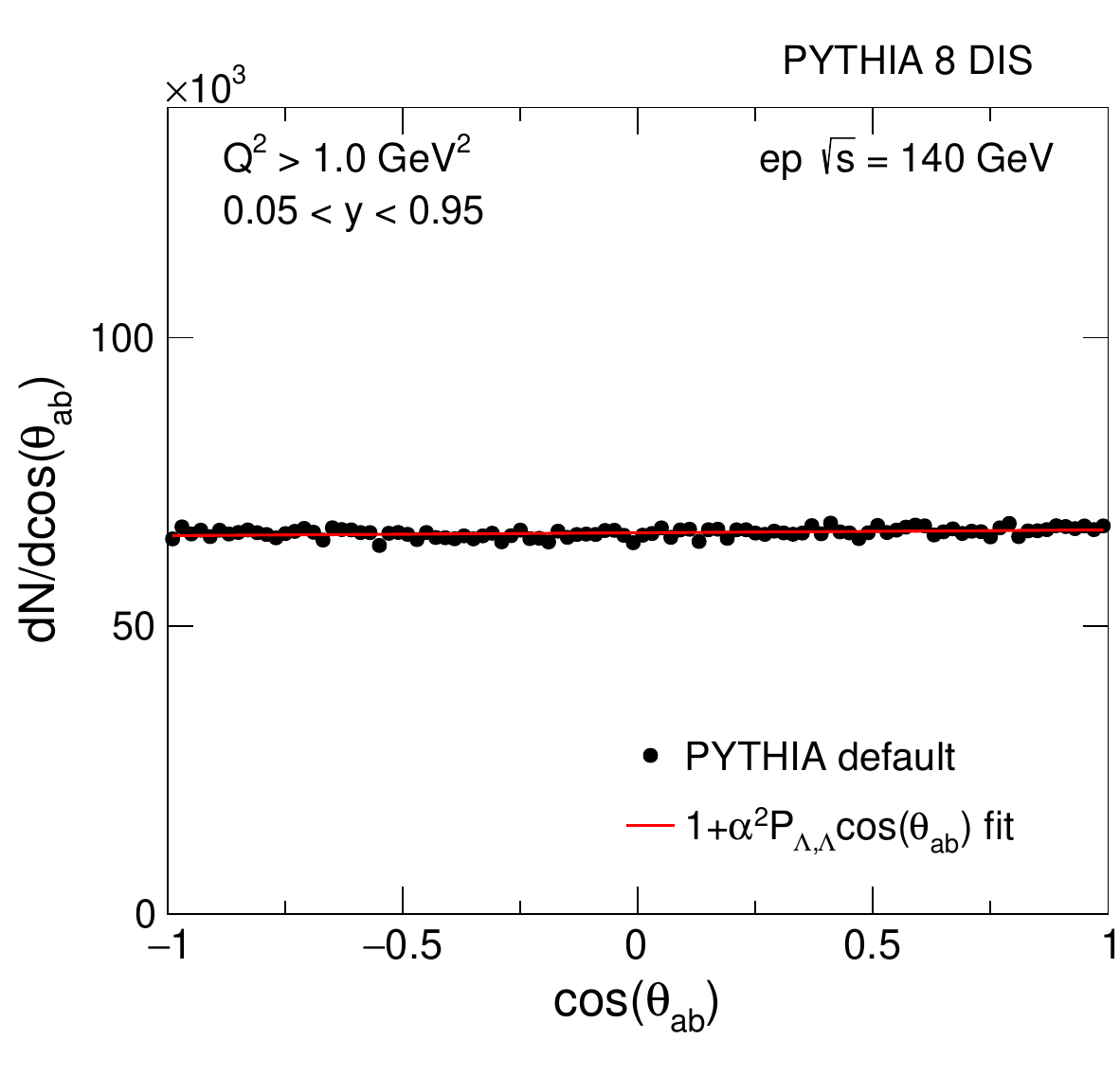}
\includegraphics[width=0.8\linewidth]{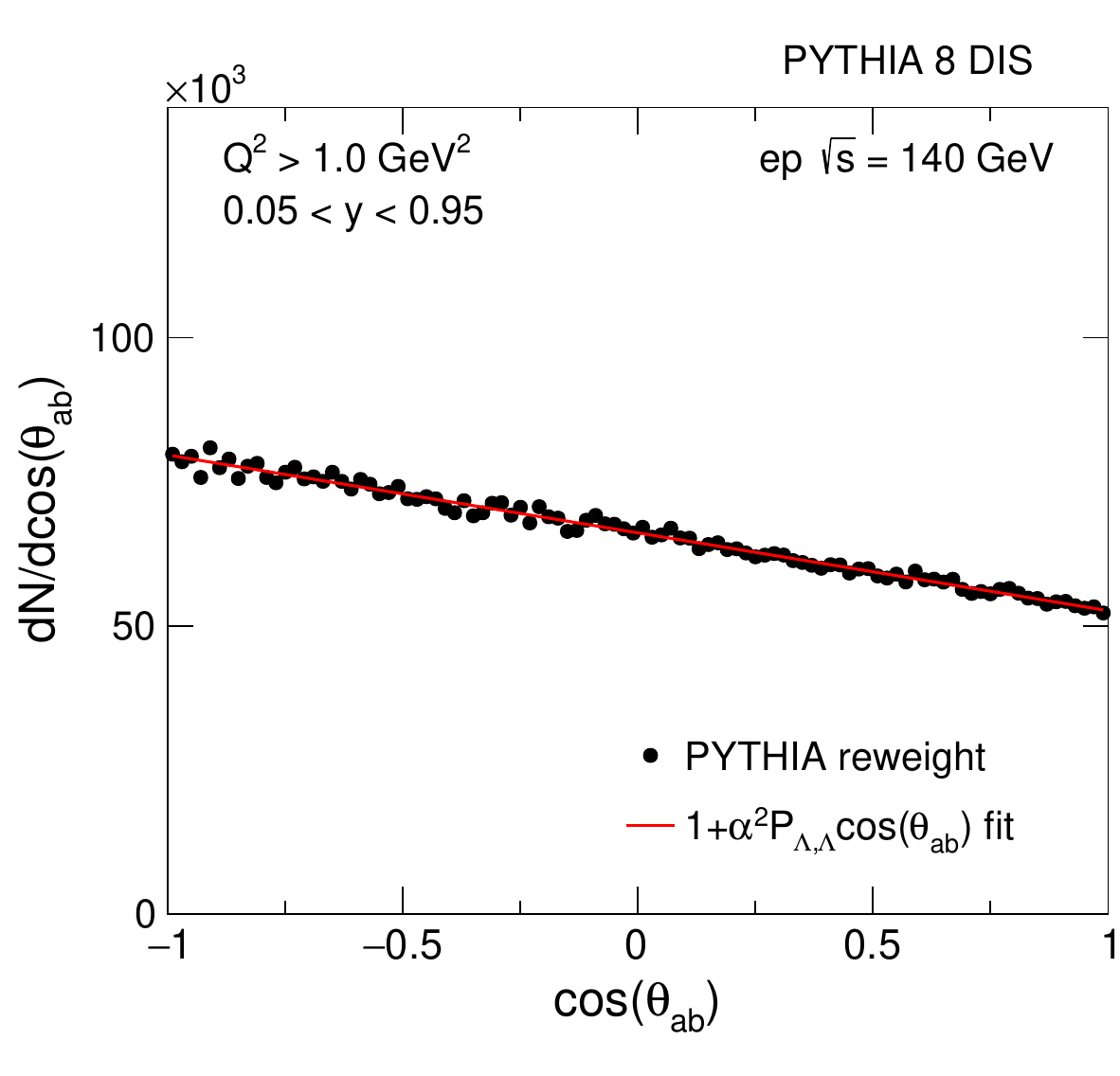}
  \caption{  Upper: the $\cos{(\theta_{\rm ab})}$ distribution is plotted for one $\Lambda$ with respect to the second $\Lambda$ particle in default PYTHIA 8~\cite{Sjostrand:2007gs} simulation. Lower: the same quantity with weighted events, where weights are determined based on polarization value $P_{\Lambda,\Lambda}=50\%$. }
  \label{fig:figure_3_appendix}
\end{figure}

In Fig.~\ref{fig:figure_3_appendix} upper, the default PYTHIA simulations are shown in terms of the cosine of the angular distribution, $\cos{(\theta_{\rm ab})}$. Here we set the parameter $n$ to be unity for simplicity. It is expected that there is no cosine modulation of the angular distribution, implying no $\Lambda\Lambda$ (or $\Lambda\bar{\Lambda}$) spin correlations. However, in Fig.~\ref{fig:figure_3_appendix} lower, we manually introduce a cosine modulation at the generator level using an event weight to show the feasibility of this measurement. Note that the nature of this cosine modulation is expected only from quantum correlations. Here the event weight is $P_{\Lambda,\Lambda}=50\%$, the angular asymmetry is found to be visible. The fitted value of $P_{\Lambda,\Lambda}$ is consistent with 50\% as was put in. The exercise of validating this observable in the PYTHIA event generator is to clearly setup the experimental baseline for such measurement, where no trivial kinematic effect can generate a background in the correlation. 

A discussion of  the measurement of $\Lambda {\bar \Lambda}$ correlations in high energy experiments would be remiss without discussing the machine capabilities and experimental challenges. Firstly, the RHIC accelerator (and the upcoming EIC)  have the capability of polarizing the target proton or light ions. By polarizing the target, the constituent quarks and gluons can be polarized such that the resulting QCD strings are in different configurations. This is an advantageous experimental handle to control the system, where an asymmetry measurement related to the $\Lambda {\bar \Lambda}$ correlation could be developed. In addition, the target polarization may provide a specific initial condition for considering the entangled Bell-states and the spin Hamiltonian, which helps data-model comparisons. 

In terms of the experiment, the rapidity separation of the two $\Lambda$ particles is an important quantity for selecting $\Lambda$s from different distant scales in a QCD string. Therefore it is essential to have a wide rapidity coverage for the experiment, where widely separated $\Lambda$ particles can be reconstructed. Without  good rapidity coverage, acceptance effects could be difficult and challenging to correct for. Finally, the measurement of $\Lambda {\bar \Lambda}$ correlations is not only constrained by the $\Lambda$'s decay properties, but also detector effects. Given the anticipated small magnitude for the correlation signal, the momentum resolution of daughter particles and the $\Lambda$ particle will significantly impact the precision of the spin projection. Therefore detectors with high momentum resolution are desirable.

\bibliography{reference}

\end{document}